\documentclass[12pt]{article}

\usepackage{verbatim}
\usepackage[utf8]{inputenc}
\usepackage{amsmath}
\DeclareMathOperator{\Tr}{Tr}

\usepackage{bbold}
\usepackage[usenames,dvipsnames,svgnames,table]{xcolor}
\definecolor{fiolet}{cmyk}{0.47,1,0.07,0.33}
\usepackage[a4paper, hmargin=2cm, tmargin=1.5cm, bmargin=3cm, footskip=2cm]{geometry}     
\usepackage{mathtools}
\usepackage{float}
\usepackage[english]{babel}
\usepackage{dsfont}
\usepackage{enumerate}
\usepackage{amssymb}
\usepackage{amsfonts}
\usepackage{amsthm}
\usepackage{bm}
\usepackage{physics}
\usepackage{tensor}
\usepackage{dsfont}
\usepackage{hyperref}
\usepackage{authblk} 
\usepackage{mathtools}
\DeclarePairedDelimiter\sbra{[}{\rvert}
\DeclarePairedDelimiter\sket{\lvert}{]}
\DeclarePairedDelimiterX\sbraket[2]{[}{]}{#1\,\delimsize\vert\,\mathopen{}#2}
\DeclarePairedDelimiterX\sabraket[2]{[}{\rangle}{#1\,\delimsize\vert\,\mathopen{}#2}
\DeclarePairedDelimiterX\asbraket[2]{\langle}{]}{#1\,\delimsize\vert\,\mathopen{}#2}
\DeclarePairedDelimiterX\smel[3]{[}{]}{#1\,\delimsize\vert\,\mathopen{}#2\,\delimsize\vert\,\mathopen{}#3}
\DeclarePairedDelimiterX\samel[3]{[}{\rangle}{#1\,\delimsize\vert\,\mathopen{}#2\,\delimsize\vert\,\mathopen{}#3}
\DeclarePairedDelimiterX\asmel[3]{\langle}{]}{#1\,\delimsize\vert\,\mathopen{}#2\,\delimsize\vert\,\mathopen{}#3}

\newcommand{\id}{\mathbb{1}}

\newcommand{\myvec}[1]{\ensuremath{\begin{pmatrix}#1\end{pmatrix}}}

\usepackage{tikz}
\usetikzlibrary{decorations.pathmorphing}
\usetikzlibrary{arrows}

\begin{document}

\counterwithin{equation}{section}

\title{Spinor - helicity calculation of the $g^* g^* \to q \overline q V^*$ amplitude at the tree level}

\author[1]{Jan Ferdyan}
\author[2]{Błażej Ruba}

\affil[1]{Institute of Theoretical Physics, Jagiellonian University, \protect\\
prof. Lojasiewicza 11, 30-348 Kraków, Poland, \protect\\
email: jan.ferdyan@student.uj.edu.pl}
\affil[2]{Department of Mathematical Sciences, University of Copenhagen, \protect\\
Universitetsparken 5, 2100 Copenhagen, Denmark, \protect\\
email: btr@math.ku.dk
}
\date{\today}
\maketitle

\abstract{We compute amplitudes for the process $g^* g^* \to q \overline q V^*$ (two virtual gluons into a~quark, antiquark and a boson) at the tree level using the spinor - helicity formalism. The resulting analytic expressions are much shorter than squared amplitudes obtained using trace methods. Our results can be used to expedite numerical calculations in phenomenological studies of the Drell -- Yan process in high energy factorization framework.
}
\section{Introduction}

The Drell -- Yan process \cite{Drell:1970wh} is a good probe for an internal structure of hadrons in proton - proton or proton - antiproton collisions. In this process a pair of lepton and anti-lepton is produced by an electroweak boson -- a virtual photon $\gamma^*$ or $Z^0$. The measured dilepton distributions can be used to determine the Drell -- Yan structure functions, see e.g. \cite{Drell:1970wh}. On the other hand, the structure functions can be predicted within QCD description based on factorization schemes: collinear or high energy factorization (also referred to as $k_T$ factorization) \cite{Collins:1991ty, Catani:1994sq, Catani:1990eg, Catani:1990xk}. The necessary input to these descriptions are parton distribution functions (PDFs), which parameterize the details of the proton structure. For the collinear factorization one applies collinear parton distribution functions, which are functions of the parton energy and of the factorization scale. In the case of high energy factorization one uses transverse momentum distributions (TMDs) \cite{Angeles-Martinez:2015sea}, which depend also on the parton transverse momentum $k_T$.

Recent measurements of the Drell -- Yan process in the $Z^0$ mass peak region at the LHC \cite{ATLAS:2016rnf} exhibit a deviation from theoretical predictions of perturbative QCD \cite{Gehrmann-DeRidder:2015wbt,Gehrmann-DeRidder:2016cdi,Gauld:2017tww,Li:2024iyj,Piloneta:2024aac} at next-to-next-to leading order (NNLO) in the Lam -- Tung combination \cite{Lam:1980uc} of the structure functions. One of the proposed explanations is that the Lam -- Tung relation breaking may occur as a result of the parton transverse momenta \cite{Peng:2015spa, Motyka:2016lta}.

In the high energy factorization, the TMDs are probed for $x \ll 1$, where $x$ is the fraction of the hadron longitudinal momentum carried by the parton. In this kinematical regime the proton structure is strongly dominated by the gluons. Following \cite{Motyka:2016lta, Deak:2008ky} we adopt the approximation in which the quark and antiquark components of the proton structure are neglected. We consider the contribution to the Drell -- Yan process which occurs by scattering of two virtual gluons $g^* g^*$ into quark, antiquark and an electroweak boson $V^*$ that decays into a dilepton $l^+l^-$. The Drell -- Yan structure functions depend on the gluon TMDs, so we can use measured data to constrain them. This requires however an efficient evaluation of the matrix elements of the process $g^* g^* \to q \bar{q} V^*$ that speeds up the fitting procedure.

We remark that the Drell-Yan process occurs also through the channel $q_{\mathrm{val}} g^* \to q  V^*$ \cite{Motyka:2016lta, Benic:2016uku, Benic:2018hvb}, involving a valence quark $q_{\mathrm{val}}$, which we do not consider in this paper. It is important when the density of valence quarks $q_{\mathrm{val}}$ is high. This is the case for $x \sim 0.1$ in the forward - backward region, in which the $V$ boson is produced almost collinearly with the incoming parton. We are most interested in central production, in which $x \ll 1$, so the contribution of valence quarks is small (they have a low parton density).

Direct evaluation of the relevant Feynman diagrams used in \cite{Motyka:2016lta} leads to very long and numerically costly expressions. In order to improve the efficiency we apply spinor - helicity formalism. The use of spinors in the study of scattering amplitudes is by now standard, see e.g.\ \cite{Kleiss:1985yh,Arkani-Hamed:2017jhn}. We use notation for spinors which is a small adaptation of the one commonly used in general relativity \cite{Penrose}.

Similar scattering processes were considered in \cite{vanHameren:2012if}, where authors studied also the possibility of multi-gluon production and performed numerical calculations. 
The amplitudes were calculated in terms of pure spinor contractions, based on the techniques from \cite{Kleiss:1985yh}. However, explicit formulas for the amplitudes directly in terms of momenta of the scattering particles were not provided. Such expressions, presented in a compact analytic form, are the main result of our work. We also present example calculations, illustrating the methods we used to obtain our results.

The $g^*g^* \to q\overline{q}V^*$ process was also studied in \cite{Motyka:2016lta, Deak:2008ky, Baranov:2008hj}, where the amplitudes squared were calculated numerically applying the standard trace approach within the $k_T$ factorization framework. Such amplitudes can be also derived within the Color Glass Condensate (CGC) framework \cite{Benic:2016uku,Benic:2018hvb} by considering the low gluon density limit. 



\subsection{Kinematics}

\begin{figure}[ht]
    \centering
    \includegraphics[height=8.8cm]{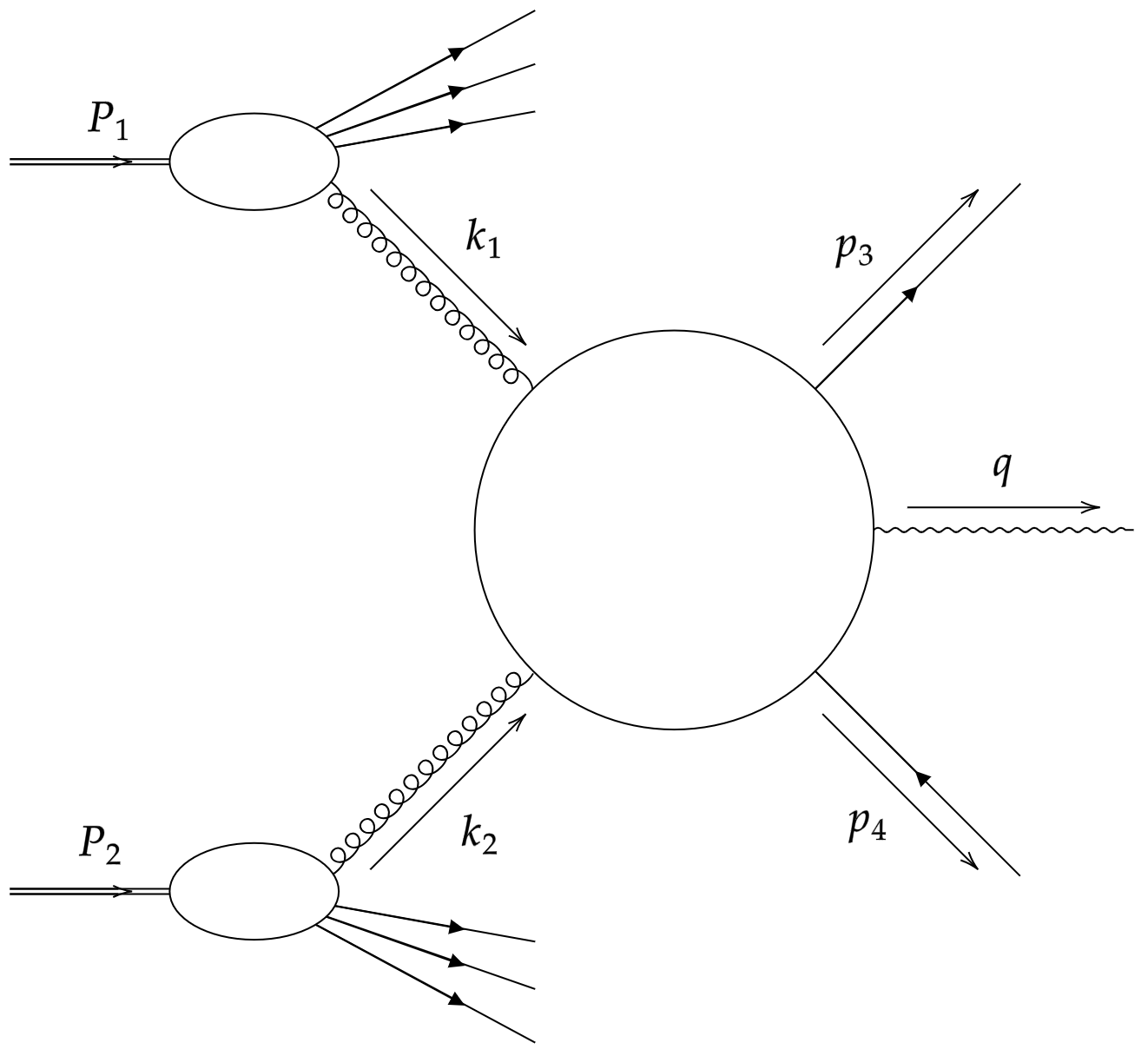}
    \caption{General form of the diagrams. Big circle represents all possible QCD LO subdiagrams. $P_1,P_2$ are the momenta of the incoming protons, $k_1,k_2$ are momenta of virtual gluons emitted by the corresponding protons, $p_3,p_4$ are the momenta of the quark and antiquark, and $q$ is the momentum of the emitted electroweak boson.}
    \label{fig:BlobDiagram}
\end{figure}

We consider production of a virtual electroweak boson in a high energy proton-proton collision. We are interested in the process illustrated on Figure \ref{fig:BlobDiagram}. The protons momenta $P_1$ and $P_2$ are near light-like, and in the center of mass system can be expressed in Minkowskian coordinates as 
\begin{equation}
 P_1 \approx \left( \sqrt{S}/2, 0, 0, \sqrt{S}/2 \right), \qquad  P_2 \approx \left( \sqrt{S}/2, 0, 0, -\sqrt{S}/2 \right).  
\end{equation}
The collision invariant energy squared is $S = \left( P_1 + P_2 \right)^2$. We work in the $k_T$ factorization framework, so partons carry nonzero transverse momenta. The full momentum $k_i$ of the gluon originating from the $i$th proton can be decomposed as
\begin{equation}
    k_i = x_i P_i + k_{iT}, \qquad i = 1,2.
\end{equation}
Here $x_i$ is the fraction of $P_i$ carried by the gluon and $k_{iT}$ is the momentum in the plane perpendicular to the scattering plane ($k_{iT} \cdot P_1 = k_{iT} \cdot P_2=0$). The gluon polarizations are approximated as \footnote{More precisely, one uses also a Ward identity: the amplitude is orthogonal to $k_i^\mu$, so contraction with $k_{iT}$ can be replaced by contraction with $-x_i P_i$.} $\epsilon^\mu(k_i) = x_i P_i^\mu/\sqrt{-k_{iT}^2}$ \cite{Collins:1991ty, Catani:1994sq, Catani:1990eg, Catani:1990xk, Deak:2008ky}. 

We denote the momentum of the boson $V^*$ as $q$ and the momenta of outgoing quark and antiquark as $p_3$ and $p_4$, respectively.

\subsection{The method and results}

We use the spinor - helicity formalism to obtain simple analytic formulas for the Feynman diagrams describing the process $g^* g^* \to q \overline q V^*$ at the tree level.

The correctness of our calculations has been checked by comparing numerically with amplitudes squared, summed over polarizations of $q$ and $\overline q$, obtained using standard trace methods. The numerical treatment was necessary because analytic formulas for traces obtained using the algebra of $\gamma$ matrices are so lengthy that it does not seem feasible to analyze them using symbolic algebra software (let alone pen and paper).

The manuscript is organized as follows. Section \ref{sec:spinor_helicity} explains our conventions for two component spinors and bispinors. In particular in Subsections \ref{sec:massless_spinor} and \ref{sec:massive_spinors} we construct polarization spinors satisfying massless or massive Dirac equation. These choices of spinor bases do not affect physical observables, such as cross sections, but they do affect the amplitudes\footnote{Simply put, amplitudes are tensors in the space of polarizations and one has to specify in which basis are theirs components given.}. In Subsection \ref{sec:components} we provide explicit expressions in terms of components for various spinorial objects, e.g.\ the dictionary between spinor components and space-time components of a vector. The main results are in Section \ref{sec:amplitudes}. We present the Feynman diagrams and the corresponding algebraic expressions, and then decompose the amplitudes according to colors and chiralities. In Subsection \ref{sec:result_massless} we evaluate the resulting expressions using spinors in the massless case. These amplitudes are less general than our final result, but due to the significant simplifications we deemed them worthy of displaying separately. In~Subsection \ref{sec:result_massive} we give formulas for amplitudes allowing for nonzero masses of the quark and the antiquark. Even though the amplitudes are significantly more complicated than in the massless case, they are still multiple orders of magnitude shorter than outputs of calculations using trace methods. In Subsection \ref{sec:calc} we present example calculations of amplitudes for some specific diagrams. We comment on possible directions for further research in Section \ref{sec:outlook}.

\section{Spinor - helicity formalism} \label{sec:spinor_helicity}

\subsection{Spinor indices and bases} \label{sec:ind_bases}

We denote (two component) spinor indices with uppercase latin indices, with an overdot used for indices in the 
complex conjugate representation. Invariant tensors $\varepsilon_{AB}, \varepsilon_{\dot A \dot B}$ (related to each other by complex conjugation) are used for raising and lowering of indices:
\begin{subequations}
\begin{gather}
\varepsilon_{AB} \varepsilon^{CB} = \tensor{\delta}{_A^C}, \label{eq:epsilon_delta} \\
\xi_B = \xi^A \varepsilon_{AB}, \\
\xi^A = \varepsilon^{AB} \xi_B,
\end{gather}
\end{subequations}
and analogously for conjugate spinor indices. The order of indices in the above formulas matters because $\varepsilon_{AB}, \varepsilon_{\dot A \dot B}$ are skew-symmetric. Thus for example
\begin{equation}
 \xi^A \eta_A = -\xi_A \eta^A .  
\end{equation}

A pair $A \dot A$ of spinor indices may be traded for one Lorentz index. This is achieved by contracting with $(\sigma^\mu)_{A \dot A}, (\overline \sigma^\mu)^{\dot A A}$, which are vectors of spinorial matrices:
\begin{equation}
v_{A \dot A} = v_\mu (\sigma^\mu)_{A \dot A}, \qquad v^{\dot A A} = v_\mu (\overline \sigma^\mu)^{\dot A A}.
\label{eq:vec_spin_trans}
\end{equation}
$\overline \sigma$ is related to $\sigma$ by raising of spinor indices combined with matrix transposition. Given two Lorentz vectors, we can contract them directly or contract their spinor indices. The results differ by a factor of $2$:
\begin{equation}
    u_{A \dot A} v^{\dot A A} = 2 u_\mu v^\mu. 
\end{equation}
Another important property of $\sigma$ and $\overline{\sigma}$ is that their product, symmetrized in Lorentz indices gives Minkowski metric
\begin{equation}
    \sigma^{\mu} \overline{\sigma}^{\nu} + \sigma^{\nu} \overline{\sigma}^{\mu}  =  2 g^{\mu \nu} \id \quad \implies \quad v_{A \dot A} v^{\dot A B} = v_\mu v^{\mu} \delta^B_A.
\end{equation}

With the aid of the inner product, we can readily obtain the general representation of a~spinor in terms of components. We choose any pair of spinors $o$ and $\iota$ normalized as
\begin{equation}
    o_A \iota^A = 1 .
\end{equation}
Then $\{ o , \iota \}$ form a basis of the spinor space. The decomposition of a general spinor $\xi$ in this basis reads
\begin{equation}
\xi = \xi^0 o + \xi^1 \iota, \qquad  \text{where} \quad  \xi^0 = \xi_A \iota^A, \quad \xi^1 = -\xi_A o^A,
\end{equation}
and the invariant tensors decompose as
\begin{equation}
    \varepsilon_{AB} = o_A \iota_B - \iota_A o_B, \qquad \varepsilon_{\dot A \dot B }= \overline o_{\dot A} \overline \iota_{\dot B} - \overline \iota_{\dot A} \overline o_{\dot B}.
\end{equation}

Now we define a null tetrad of world-vectors as 
\begin{equation}
\label{null basis}
\begin{split}
    l_+^{\dot{A} A} = o^A \overline{o}^{\dot{A}} \qquad &\iff \qquad l_+^\mu = \frac{1}{2} o^A \sigma^\mu_{A \dot{A}} \overline{o}^{\dot{A}}, \\
    l_-^{\dot{A} A} = \iota^A \overline{\iota}^{\dot{A}} \qquad &\iff \qquad l_-^\mu = \frac{1}{2} \iota^A \sigma^\mu_{A \dot{A}} \overline{\iota}^{\dot{A}}, \\
    m^{\dot{A} A} = o^A \overline{\iota}^{\dot{A}} \qquad &\iff \qquad m^\mu = \frac{1}{2} o^A \sigma^\mu_{A \dot{A}} \overline{\iota}^{\dot{A}}, \\
    \overline{m}^{\dot{A} A} = \iota^A \overline{o}^{\dot{A}} \qquad &\iff \qquad \overline{m}^\mu = \frac{1}{2} \iota^A \sigma^\mu_{A \dot{A}} \overline{o}^{\dot{A}}.
\end{split}
\end{equation}
Using the above null basis we can construct a standard orthonormal basis for Minkowski space - a Minkowski tetrad $\{t^\mu, x^\mu, y^\mu, z^\mu \}$:
\begin{equation}
\begin{split}
    l_\pm = t \pm z \quad \iff& \quad t = \frac{1}{2} \left( l_+ + l_- \right), \quad z = \frac{1}{2} \left( l_+ - l_- \right), \\
    m = x + iy, \quad \overline{m} = x - iy \quad \iff& \quad x = \frac{1}{2} \left( m + \overline{m} \right), \quad y = \frac{1}{2i} \left( m - \overline{m} \right).
\end{split}
\end{equation}
In principle we could start with the null tetrad or Minkowski tetrad and construct from them a~spinor basis $\{o, \iota\}$, unique up to an overall sign (the same for $o$ and $\iota$). 

\subsection{Dirac bispinors}


A Dirac bispinor is a pair of two spinors:
\begin{equation}
    \psi = \begin{bmatrix}
        \xi_A \\ \overline{\eta}^{\dot{A}}
    \end{bmatrix}, \qquad \overline{\psi} = \begin{bmatrix}
        \eta^A & \overline{\xi}_{\dot{A}}
    \end{bmatrix},
\end{equation}
where $\overline{\psi}$ is the Dirac conjugate of $\psi$. It is clear from the above definition that the identity $\overline{\psi_1}\psi_2 = \overline{\overline{\psi_2}\psi_1}$ holds. We work with a chiral representation of gamma matrices
\begin{equation}
    \gamma^\mu = \begin{bmatrix}
        0 & \sigma^\mu \\
        \overline{\sigma}^\mu & 0
    \end{bmatrix}, \quad \quad \gamma_5 = \begin{bmatrix}
        -\id & 0 \\
        0 & \id
    \end{bmatrix}.
\end{equation}
We will also use the eigenprojections of $\gamma_5$:
\begin{equation}
    P_{\pm} = \frac{1 \pm \gamma_5}{2}. 
\end{equation}

For any Lorentz vector $n$ we define the matrix
\begin{equation}
    \widehat{n} = n_\mu \gamma^\mu = \begin{bmatrix} 0 & n_\mu \sigma^\mu \\ n_\mu \overline{\sigma}^\mu & 0 \end{bmatrix} = \begin{bmatrix} 0 & n_{A\dot{A}} \\ n^{\dot{A}A} & 0 \end{bmatrix}.
\end{equation}

\subsection{Spinor bracket notation} \label{sec:massless_spinor}

We will now construct a convenient basis of bispinors satisfying the massless Dirac equation.

We choose a reference spinor $\rho^A \neq 0$ and define a future-directed null vector $l$ by $l^{A \dot A} = \rho^A \overline \rho^{\dot A}$. For every null, future directed, real Lorentz vector $n$ with $n \cdot l \neq 0$ we introduce the angle and square bra and ket spinors:
\begin{equation}
\label{brackets definition}
    \sket{n} = \begin{bmatrix} \theta(n)_A \\ 0 \end{bmatrix}, \quad \ket{n} = \begin{bmatrix} 0 \\ \overline{\theta(n)}^{\dot{A}} \end{bmatrix}, \quad \sbra{n} = \begin{bmatrix} \theta(n)^A & 0 \end{bmatrix}, \quad \bra{n} = \begin{bmatrix} 0 & \overline{\theta(n)}_{\dot{A}} \end{bmatrix},
\end{equation}
where $\theta(n)_A = \frac{1}{\sqrt{2\abs{n \cdot l}}} n_{A \dot A} \overline \rho^{\dot A}$. Then $\sket{n}, \ket{n}$, resp. $\sbra{n}, \bra{n}$ form bases of solutions of 
\begin{equation}
    \widehat n \psi =0 , \quad \text{resp.} \quad \overline \psi \widehat n =0.
\end{equation}
The normalization factor $\frac{1}{\sqrt{2 \abs{n \cdot l}}}$ is needed for the normalization of the current:
\begin{equation}
    \overline \psi \gamma^\mu \psi = 2 n^\mu.
    \label{eq:spinor_norm}
\end{equation}
We remark that the factor $\frac{1}{\sqrt{2\abs{n \cdot l}}}$ makes our solutions singular when $n$ is colinear with $l$. This is just a singularity of the chosen basis and depends on the choice of the reference spinor.

The spinors defined in \eqref{brackets definition} are eigenvectors of $\gamma_5$:
\begin{equation}
    \gamma_5 \sket{n} = -\sket{n}, \quad \gamma_5 \ket{n} = \ket{n}, \quad \sbra{n} \gamma_5 = -\sbra{n}, \quad \bra{n} \gamma_5 = \bra{n},
    \label{eq:spinor_chirality}
\end{equation}
so the basis vectors defined in \eqref{brackets definition} correspond to particles of fixed chirality (and hence also helicity). Spinors of different helicity are orthogonal, e.g.\ we have $\asbraket{n_1}{n_2}=0$ and (since $\gamma^\mu$ flips chirality) $\sbra{n_1} \gamma^\mu  \sket{n_2} =0$.

\subsection{Massive Dirac bispinors} \label{sec:massive_spinors}

For a Lorentz vector $p$ with $p^2 = m^2>0 $ we define the brackets as
\begin{equation}
\label{massive brackets definition}
    \sket{p} = \begin{bmatrix} \theta(p)_A \\ \overline{\phi(p)}^{\dot{A}} \end{bmatrix}, \quad \ket{p} = \begin{bmatrix} -\phi(p)_A \\ \overline{\theta(p)}^{\dot{A}} \end{bmatrix}, \quad \sbra{p} = \begin{bmatrix} \theta(p)^A & -\overline{\phi(p)}_{\dot{A}} \end{bmatrix}, \quad \bra{p} = \begin{bmatrix} \phi(p)^A & \overline{\theta(p)}_{\dot{A}} \end{bmatrix},
\end{equation}
where
\begin{equation}
\label{massive composite spinors}
    \theta(p)_A \coloneqq \frac{p_{A\dot{A}} \overline{\rho}^{\dot{A}}}{\sqrt{2|p \cdot l|}}, \qquad \phi(p)_A \coloneqq \frac{m}{\sqrt{2|p \cdot l|}} \rho_A.
\end{equation}
This reduces to the definition in \eqref{brackets definition} for $m \to 0$, so the notation is consistent.

The constructed spinors satisfy the Dirac equation
\begin{equation}
    (\widehat{p} - m) \sket{p} = (\widehat{p} - m) \ket{p} = 0 = \sbra{p} (\widehat{p} - m) = \bra{p} (\widehat{p} - m).
\end{equation}
Unlike in the massless case, spinor brackets have no singularities because $p \cdot l \neq 0$ for all vectors $p$ with $p^2 > 0 $. Another difference is that the direction of their spin depends on the reference spinor. More precisely, we have eigenequations for the spinor operator $S_p = \frac{1}{2 p \cdot l} \gamma_5 [\widehat p , \widehat l] $:
\begin{equation}
S_p |p ] = - | p], \quad S_p | p \rangle = | p \rangle, \quad [ p | S_p = - [ p |, \quad  \langle p | S_p = \langle p |.
\label{eq:spin_eigenequation}
\end{equation}

We describe a particle with momentum $p$ (a future-directed vector) with kets $\ket{p}$ or $\sket{p}$, depending on the value of the spin. For an anti-particle we use instead the spinors $\ket{-p}, \sket{-p}$ (note that definitions in \eqref{massive brackets definition} and \eqref{massive composite spinors} make sense both for future and past-directed vectors).

We will now prove useful symmetries of spinor brackets, which allow to reduce the number of amplitudes that we will need to compute by a factor $2$. Let us consider the charge conjugation $\mathcal{C}$ which acts on bispinors in the following way
\begin{equation}
    \mathcal{C} \psi = \mathcal{C} \begin{bmatrix} \xi_A \\ \overline{\eta}^{\dot{A}} \end{bmatrix} = \begin{bmatrix} \eta_A \\ \overline{\xi}^{\dot{A}} \end{bmatrix} .
\end{equation}
The operator $\mathcal{C}$ is anti-linear, $\mathcal{C}^2 = \id$ and it anti-commutes with gamma matrices $\gamma^\mu$ and $\gamma_5$.
Now we can define another operator
\begin{equation}
    \mathcal{Q} \coloneqq \gamma_5 \mathcal{C},
\end{equation}
which satisfies 
\begin{equation}
    \mathcal{Q}^2 = -\id, \qquad \mathcal{Q} \gamma_5 \mathcal{Q}^{-1} = -\gamma_5, \qquad \mathcal{Q} \gamma^\mu \mathcal{Q}^{-1} = \gamma^\mu.
\end{equation}
We also have, for any bispinors $\psi_1, \psi_2$:
\begin{equation}
    \overline{\mathcal{Q} \psi_1} \mathcal{Q} \psi_2 = \overline{\overline{\psi_1} \psi_2}.
\end{equation}
It can be also directly shown that for bispinor brackets we have the following identities
\begin{equation}
    \mathcal{Q} \sket{p} = \ket{p}, \qquad \mathcal{Q} \ket{p} = -\sket{p}.
\end{equation}

With the help of the above identities we will derive useful symmetries. Let us begin with the contraction with $\Gamma$ being any composition of the gamma matrices
\begin{equation}
\label{symmetry1}
    \samel{p_3}{\Gamma}{p_4} = \overline{\ket{p_3}} \Gamma \ket{p_4} = \overline{\mathcal{Q}\sket{p_3}} \Gamma \mathcal{Q} \sket{p_4} = \overline{\mathcal{Q}\sket{p_3}} \mathcal{Q} \Gamma \sket{p_4} = \overline{\overline{\sket{p_3}} \Gamma \sket{p_4}} = \overline{\asmel{p_3}{\Gamma}{p_4}},
\end{equation}
and similarly
\begin{equation}
\label{symmetry2}
    \smel{p_3}{\Gamma}{p_4} = -\overline{\mathcal{Q}\sket{p_3}} \Gamma \mathcal{Q} \ket{p_4} = -\overline{\mathcal{Q}\sket{p_3}} \mathcal{Q} \Gamma \ket{p_4} = -\overline{\overline{\sket{p_3}} \Gamma \ket{p_4}} = -\overline{\mel{p_3}{\Gamma}{p_4}}.
\end{equation}
These symmetries hold for both the massless and massive case.

Because $\mathcal{Q}$ anticommutes with $\gamma_5$, it exchanges the projections $\mathcal{Q} P_\pm \mathcal{Q}^{-1} = P_\mp$. Therefore if we now define $\Gamma_\pm$ to be a composition of the gamma matrices and one of the projections $P_\pm$, the symmetries \eqref{symmetry1}, \eqref{symmetry2} translates to
\begin{equation}
\label{symmetry3}
    \samel{p_3}{\Gamma_\pm}{p_4} = \overline{\mathcal{Q}\sket{p_3}} \Gamma_\pm \mathcal{Q} \sket{p_4} = \overline{\mathcal{Q}\sket{p_3}} \mathcal{Q} \Gamma_\mp \sket{p_4} = \overline{\asmel{p_3}{\Gamma_\mp}{p_4}},
\end{equation}
and
\begin{equation}
\label{symmetry4}
    \smel{p_3}{\Gamma_\pm}{p_4} = -\overline{\mathcal{Q}\sket{p_3}} \mathcal{Q} \Gamma_\mp \ket{p_4} = -\overline{\mel{p_3}{\Gamma_\mp}{p_4}}.
\end{equation}


\subsection{Component expressions} \label{sec:components}

We will work with null tetrad defined as in Subsection \ref{sec:ind_bases}. Vectors can be written with components with respect to chosen basis as
\begin{equation}
    n^\mu = n^+ l_+^\mu + n^- l_-^\mu + n^\perp m^\mu + n^{\overline{\perp}} \overline{m}^\mu = \begin{pmatrix}
        n^+ & n^- & n^\perp & n^{\overline{\perp}}
    \end{pmatrix}^\mu,
\end{equation}
where
\begin{equation}
    n^\pm = n^0 \pm n^3, \quad n^\perp = n^1 + in^2, \quad n^{\overline{\perp}} = n^1 - in^2.
\end{equation}
The correspondence with spinor components is:
\begin{equation}
    n^{\dot A A} = \begin{pmatrix}
        n^{\dot 0 0} & n^{\dot 0 1} \\ n^{\dot 1 0} & n^{\dot 1 1}
    \end{pmatrix} = \begin{pmatrix}
        n^+ & n^{\overline{\perp}} \\ n^\perp & n^-
    \end{pmatrix}.
\end{equation}
The Minkowski metric in the null basis has the form
\begin{equation}
    \left( g_{\mu \nu} \right) = \frac{1}{2} \myvec{ 0 & 1 & 0 & 0 \\
    1 & 0 & 0 & 0 \\
    0 & 0 & 0 & -1 \\
    0 & 0 & -1 & 0 }, \quad \quad \left( g^{\mu \nu} \right) = 2 \myvec{ 0 & 1 & 0 & 0 \\
    1 & 0 & 0 & 0 \\
    0 & 0 & 0 & -1 \\
    0 & 0 & -1 & 0 }.
\end{equation}
We note that for real vectors $n^{\overline{\perp}} = \overline{n^{\perp}}$, but for complex vectors we have $n^{\overline{\perp}} = \overline{\overline{n}^{\perp}}$.

\begin{equation}
    \overline{n}^\mu = \begin{pmatrix}
        \overline{n}^+ & \overline{n}^- & \overline{n}^\perp & \overline{n}^{\overline{\perp}}
    \end{pmatrix}^\mu = \begin{pmatrix}
        \overline{n^+} & \overline{n^-} & \overline{n^{\overline{\perp}}} & \overline{n^\perp}
    \end{pmatrix}^\mu
\end{equation}

We choose the basis spinors to be
\begin{equation}
    o_A = \myvec{0 \\ 1}, \quad \iota_A = \myvec{-1 \\ 0}, \quad o^A = \varepsilon^{AB} o_B = \myvec{1 \\ 0}, \quad \iota^A = \varepsilon^{AB} \iota_B = \myvec{0 \\ 1},
\end{equation}
thus the symplectic form in the matrix form is
\begin{equation}
    \varepsilon_{AB} = o_A \iota_B - \iota_A o_B = \myvec{0 & 1 \\ -1 & 0} = \varepsilon^{AB} = o^A \iota^B - \iota^A o^B.
\end{equation}
With this choice of spinor basis we define the bispinor basis
\begin{equation}
\label{bispinor basis}
\begin{split}
    \sket{\uparrow} = \begin{bmatrix} o_A \\ 0 \end{bmatrix} = \begin{bmatrix} 0 \\ 1 \\ 0 \\ 0 \end{bmatrix}, \quad \ket{\uparrow} = \begin{bmatrix} 0 \\ \overline{o}^{\dot{A}} \end{bmatrix} = \begin{bmatrix} 0 \\ 0 \\ 1 \\ 0 \end{bmatrix},& \quad \sket{\downarrow} = \begin{bmatrix} \iota_A \\ 0 \end{bmatrix} = \begin{bmatrix} -1 \\ 0 \\ 0 \\ 0 \end{bmatrix}, \quad \ket{\downarrow} = \begin{bmatrix} 0 \\ \overline{\iota}^{\dot{A}} \end{bmatrix} = \begin{bmatrix} 0 \\ 0 \\ 0 \\ 1 \end{bmatrix}, \\
    \sbra{\uparrow} = \begin{bmatrix} o^A & 0 \end{bmatrix} = \begin{bmatrix} 1 & 0 & 0 & 0 \end{bmatrix},& \quad \bra{\uparrow} = \begin{bmatrix} 0 & \overline{o}_{\dot{A}} \end{bmatrix} = \begin{bmatrix} 0 & 0 & 0 & 1 \end{bmatrix}, \\
    \sbra{\downarrow} = \begin{bmatrix} \iota^A & 0 \end{bmatrix} = \begin{bmatrix} 0 & 1 & 0 & 0 \end{bmatrix},& \quad \bra{\downarrow} = \begin{bmatrix} 0 & \overline{\iota}_{\dot{A}} \end{bmatrix} = \begin{bmatrix} 0 & 0 & -1 & 0 \end{bmatrix}.
\end{split}
\end{equation}
From this basis one can check by explicit calculation that for the basis vectors \eqref{null basis} we have the completeness relation
\begin{subequations}
\label{completeness relation for basis}
\begin{gather}
    \widehat{l}_+ = \ket{\uparrow}\sbra{\uparrow} + \sket{\uparrow}\bra{\uparrow}, \qquad \samel{\uparrow}{\gamma^\mu}{\uparrow} = 2l_+^\mu, \\
    \widehat{l}_- = \ket{\downarrow}\sbra{\downarrow} + \sket{\downarrow}\bra{\downarrow}, \qquad \samel{\downarrow}{\gamma^\mu}{\downarrow} = 2l_-^\mu, \\
    \widehat{m} = \ket{\downarrow}\sbra{\uparrow} + \sket{\uparrow}\bra{\downarrow}, \qquad \samel{\uparrow}{\gamma^\mu}{\downarrow} = 2m^\mu, \\
    \widehat{\overline{m}} = \ket{\uparrow}\sbra{\downarrow} + \sket{\downarrow}\bra{\uparrow}, \qquad \samel{\downarrow}{\gamma^\mu}{\uparrow} = 2\overline{m}^\mu.
\end{gather}
\end{subequations}
Using the bispinor basis we can write also the completeness relations in eigenspaces of chirality:
\begin{equation}
\label{Completeness relation}
    \quad P_\pm = \frac{\id \pm \gamma_5}{2}, \quad P_+ = \ket{\downarrow}\bra{\uparrow} - \ket{\uparrow}\bra{\downarrow}, \quad P_- = \sket{\uparrow}\sbra{\downarrow} - \sket{\downarrow}\sbra{\uparrow}.
\end{equation}

Let us choose the reference spinor $\rho^A$ to be $\iota^A$. Then we have $l = l_-$ and $2(p \cdot l) = p^+$. The Weyl spinors $\theta(p)_A$ used to construct solutions of the Dirac equation take the form: 
\begin{equation}
    \theta(p)_A = \frac{p_{A \dot 1}}{\sqrt{|p^+|}} = \frac{1}{\sqrt{|p^+|}} \left( p^+ o_A + p^{\overline{\perp}} \iota_A \right) = \frac{1}{\sqrt{|p^+|}} \begin{pmatrix}
        -p^{\overline{\perp}} \\ p^+
    \end{pmatrix}.
\end{equation}
Therefore the corresponding bispinor brackets are
\begin{equation}
\label{bispinors in basis}
    \sket{p} = \frac{1}{\sqrt{\abs{p^+}}} \left( p^+ \sket{\uparrow} + p^{\overline{\perp}} \sket{\downarrow} + m \ket{\downarrow} \right), \qquad \ket{p} = \frac{1}{\sqrt{\abs{p^+}}} \left( p^+ \ket{\uparrow} + p^{\perp} \ket{\downarrow} - m \sket{\downarrow} \right).
\end{equation}


\section{Amplitudes} \label{sec:amplitudes}

\subsection{Derivation of the amplitudes}

\begin{figure}[ht]
    \centering
    \includegraphics[height=9cm]{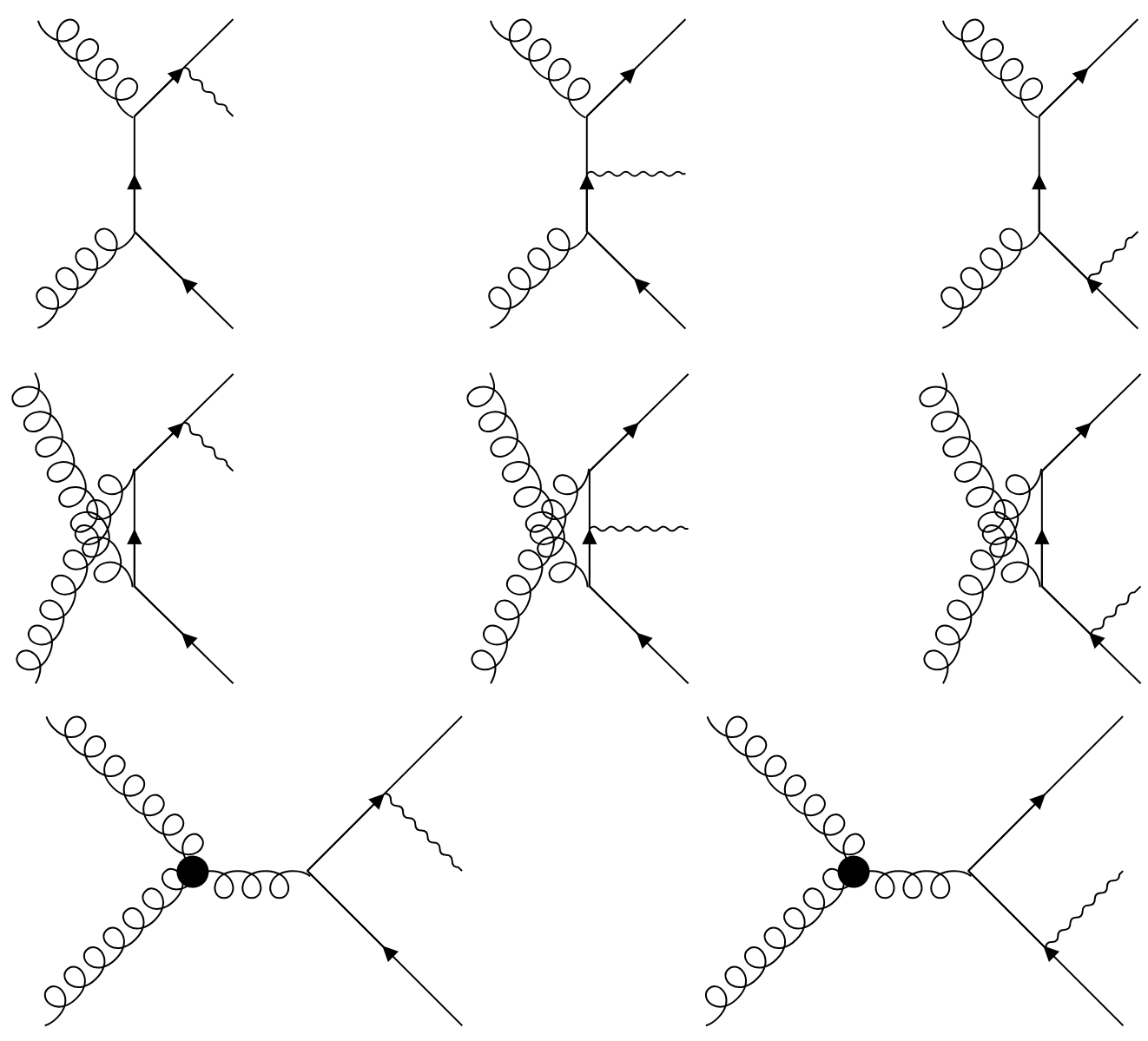}
    \caption{Eight Feynman diagrams describing the $g^* g^* \to \overline{q} q V^*$ scattering at the tree level. The bold dot represents the Lipatov vertex.}
    \label{fig:AllDiagrams}
\end{figure}

The amplitude for the $g^* g^* \to \overline{q} q V^*$ scattering at the tree level is given by diagrams presented on Figure \ref{fig:AllDiagrams}. Some care is needed to define the amplitude involving off-shell gluons in a gauge-invariant way, e.g.~by embedding as a subprocess in an on-shell process of scattering of two fast quarks \cite{Motyka:2016lta, Deak:2008ky, vanHameren:2012if}. This leads to a replacement of the ordinary QCD three - gluon vertex with the so - called Lipatov vertex \cite{Lipatov:1995pn} (defined in \eqref{eq:Lipatov_vertex} below), which, apart from the standard $ggg$ interaction in QCD, takes into account an exchange of gluon between the two quarks and gluon radiation.
%

We take the vertex of interaction with the boson $V^*$ to be
\begin{equation}
 \Gamma^\mu = v 
 \gamma^\mu + a \gamma_5 \gamma^\mu.   
\end{equation}
For example, if $V^* = \gamma^*$ (excited photon), then $v= e_f$ and $a=0$. We need the more general vertex to treat the $W$ and $Z$ bosons. We denote colors of the ingoing gluons by $a,b$. The $q \overline q$ pair is taken to have colors $i, j$, and spins $\sigma_3, \sigma_4$. The contributions to the amplitude from the eight diagrams take the following forms:
\begin{equation}
\begin{split}
    \mathcal{A}_{1}^\mu &= \frac{-i g^2  }{\left( v_1^2 - m_4^2 \right)  \left( v_2^2 - m_4^2 \right)} \left(T^a T^b\right)_{ij} \overline{u}_{\sigma_3}(p_3)  \Gamma^\mu \left(\widehat{v}_1 + m_4 \right) \widehat{P}_1 \left(\widehat{v}_2 + m_4 \right) \widehat{P}_2 v_{\sigma_4} (p_4) = \\
    &=: -i g^2 (T^a T^b)_{ij}  A_1^\mu,
\end{split}
\end{equation}
\begin{equation}
\begin{split}
    \mathcal{A}_{2}^\mu &= \frac{-i g^2}{\left( v_3^2 - m_3^2 \right) \left( v_2^2 - m_4^2 \right)} \left(T^a T^b\right)_{ij} \overline{u}_{\sigma_3}(p_3) \widehat{P}_1 \left(\widehat{v}_3 + m_3 \right) \Gamma^\mu \left(\widehat{v}_2 + m_4 \right) \widehat{P}_2 v_{\sigma_4} (p_4)  = \\
    &=: -i g^2 \left(T^a T^b\right)_{ij} A_2^\mu,
\end{split}
\end{equation}
\begin{equation}
\begin{split}
    \mathcal{A}_{3}^\mu &= \frac{-i g^2}{\left( v_3^2 - m_3^2 \right) \left( v_4^2 - m_3^2 \right)} \left(T^a T^b\right)_{ij} \overline{u}_{\sigma_3}(p_3) \widehat{P}_1 \left(\widehat{v}_3 + m_3 \right) \widehat{P}_2 \left(\widehat{v}_4 + m_3 \right) \Gamma^\mu v_{\sigma_4} (p_4) = \\
    &=: -i  g^2 \left(T^a T^b\right)_{ij} A_3^\mu,
\end{split}
\end{equation}
\begin{equation}
\begin{split}
    \mathcal{A}_{4}^\mu &= \frac{-i  g^2}{\left( v_1^2 - m_4^2 \right) \left( v_5^2 - m_4^2 \right)} \left(T^b T^a\right)_{ij} \overline{u}_{\sigma_3}(p_3) \Gamma^\mu \left(\widehat{v}_1 + m_4 \right) \widehat{P}_2 \left(\widehat{v}_5 + m_4 \right) \widehat{P}_1 v_{\sigma_4} (p_4) = \\
    &=: -i  g^2 \left(T^b T^a\right)_{ij} A_4^\mu,
\end{split}
\end{equation}
\begin{equation}
\begin{split}
    \mathcal{A}_{5}^\mu &= \frac{-i  g^2}{\left( v_6^2 - m_3^2 \right) \left( v_5^2 - m_4^2 \right)} \left(T^b T^a\right)_{ij} \overline{u}_{\sigma_3}(p_3) \widehat{P}_2 \left(\widehat{v}_6 + m_3 \right) \Gamma^\mu \left(\widehat{v}_5 + m_4 \right) \widehat{P}_1 v_{\sigma_4} (p_4) =  \\
    &=: -i  g^2 \left(T^b T^a\right)_{ij} A_5^\mu,
\end{split}
\end{equation}
\begin{equation}
\begin{split}
    \mathcal{A}_{6}^\mu &= \frac{-i  g^2}{\left( v_4^2 - m_3^2 \right) \left( v_6^2 - m_3^2 \right)} \left(T^b T^a\right)_{ij} \overline{u}_{\sigma_3}(p_3) \widehat{P}_2 \left(\widehat{v}_6 + m_3 \right) \widehat{P}_1 \left(\widehat{v}_4 + m_3 \right) \Gamma^\mu v_{\sigma_4} (p_4) = \\
    &=: -i  g^2 \left(T^b T^a\right)_{ij} A_6^\mu,
\end{split}
\end{equation}
\begin{equation}
\begin{split}
    \mathcal{A}_{7}^\mu = \frac{ g^2 }{\left( k_1 + k_2 \right)^2 \left( v_1^2 - m_4^2 \right)} f^{abc} T^c_{ij} \overline{u}_{\sigma_3}(p_3) \Gamma^\mu \left(\widehat{v}_1 + m_4 \right) \widehat{V}_{\mathrm{eff}} v_{\sigma_4} (p_4) =:  g^2 f^{abc} T^c_{ij} A_7^\mu,
\end{split}
\end{equation}
\begin{equation}
\begin{split}
    \mathcal{A}_{8}^\mu = \frac{ g^2 }{\left( k_1 + k_2 \right)^2 \left( v_4^2 - m_3^2 \right)} f^{abc} T^c_{ij} \overline{u}_{\sigma_3}(p_3) \widehat{V}_{\mathrm{eff}} \left(\widehat{v}_4 + m_3 \right) \Gamma^\mu v_{\sigma_4} (p_4) =:  g^2 f^{abc} T^c_{ij} A_8^\mu,
\end{split}
\end{equation}
where $v_i$ are momenta flowing in internal lines:
\begin{equation}
\begin{split}
    v_1 =& p_3 + q, \quad v_2 = k_2 - p_4, \quad v_3 = p_3 - k_1, \\
    v_4 =& -p_4 - q, \quad v_5 = k_1 - p_4, \quad v_6 = p_3 - k_2,
    \end{split}
\end{equation}
and the form of $V_{\mathrm{eff}}^\mu$ results from the contraction of the Lipatov vertex with the approximated gluon polarizations:
\begin{equation}
\begin{split}
\label{eq:Lipatov_vertex}
    V_{\mathrm{eff}}^\mu =& \frac{S}{2} (k_2 - k_1)^\mu + \left( 2 P_2 \cdot k_1 + \frac{P_1 \cdot P_2}{P_1 \cdot k_2} k_1^2 \right) P_1^\mu - \left( 2 P_1 \cdot k_2 + \frac{P_1 \cdot P_2}{P_2 \cdot k_1} k_2^2 \right) P_2^\mu = \\
    =& \frac{S}{2} (k_2 - k_1)^\mu + \left( x_1 S + \frac{k_1^2}{x_2} \right) P_1^\mu - \left( x_2 S + \frac{k_2^2}{x_1} \right) P_2^\mu.
        \end{split}
\end{equation}

We have factorized the amplitudes $\mathcal A_i^\mu$ into color factors and color-independent amplitudes $A_i^\mu$. The full amplitude can be decomposed into parts symmetric and antisymmetric in the color indices $a,b$:
\begin{equation}
    \mathcal{A}^\mu \coloneqq \sum_{n=1}^8 \mathcal{A}_n^\mu = \mathcal{A}_S^\mu + \mathcal{A}_A^\mu,
\end{equation}
where
\begin{equation}
\begin{split}
    \mathcal{A}_S^\mu \coloneqq& -i g^2 \left( \frac{1}{N} \delta^{ab} \delta_{ij} + d^{abc} T^c_{ij} \right) A_S^\mu, \\
    \mathcal{A}_A^\mu \coloneqq& g^2 f^{abc} T^c_{ij} A_A^\mu.
\end{split}
\end{equation}
and
\begin{equation}
    A_S^\mu \coloneqq \frac{1}{2} \sum_{n=1}^6 A_n^\mu, \quad A_A^\mu \coloneqq \frac{1}{2} \left( A_1^\mu + A_2^\mu + A_3^\mu - A_4^\mu - A_5^\mu - A_6^\mu \right) + A_7^\mu + A_8^\mu.
\end{equation}
In the amplitude squared, averaged over colors of the ingoing gluons and summed over colors of the outgoing $q \overline q$, the symmetric and antisymmetric parts do not interfere:
\begin{equation}
\label{Amplitude squared with spins}
\begin{split}
    \mathcal{M}^{\mu \nu} =& \frac{1}{\left( N^2 - 1 \right)^2} \sum_{i, j, a, b} \mathcal{A}^\mu \overline{\mathcal{A}}^{ \nu} = \\
    =& \frac{1}{\left( N^2 - 1 \right)^2} \sum_{i, j, a, b} \mathcal{A}_S^\mu \overline{\mathcal{A}}_S^{\nu} + \frac{1}{\left( N^2 - 1 \right)^2} \sum_{i, j, a, b} \mathcal{A}_A^\mu \overline{\mathcal{A}}_A^{ \nu} = \\
    = & \frac{g^4 \left( N^2-2 \right)}{2N \left( N^2 - 1 \right)} A_S^\mu \overline{A}_S^{\nu} + \frac{g^4 N}{2\left( N^2 - 1 \right)} A_A^\mu \overline{A}_A^{ \nu} = \mathcal M^{\mu \nu}_S + \mathcal M^{\mu \nu}_A.
\end{split}
\end{equation}

We can further decompose each amplitude into right $R^\mu$ and left $L^\mu$ part
\begin{equation}
    A_n^\mu = (v+a) R_n^\mu + (v-a) L_n^\mu,
\end{equation}
where in $R_i^\mu$ we replace $\Gamma^\mu$ by $P_+ \gamma^\mu$ and in $L_i^\mu$ we replace $\Gamma^\mu$ by $P_- \gamma^\mu$.

In the brackets notation we put for fermions 
\begin{equation}
    u_+(p) = \ket{p}, \qquad u_-(p) = \sket{p}, \qquad \overline{u}_+(p) = \sbra{p}, \qquad \overline{u}_-(p) = \bra{p},
\end{equation}
and for anti-fermions
\begin{equation}
    v_-(p) = \ket{-p}, \qquad v_+(p) = \sket{-p}, \qquad \overline{v}_-(p) = \sbra{-p}, \qquad \overline{v}_+(p) = \bra{-p},
\end{equation}
with brackets defined as in Section \ref{sec:spinor_helicity}. In the massless case, the indices $\pm$ refer to helicity (so they agree with chirality for the quark and are opposite to chirality for the antiquark). For massive particles, $\pm \frac{1}{2}$ is minus the spin in rest frame of $p$ projected onto the spatial direction of $l_-$. Equivalently, they are eigenvectors of the spin operator $S_p$ as described in \ref{eq:spin_eigenequation} (where we have to keep in mind that for antiparticles the physical spin is opposite to the spin operator from Dirac's theory).

\subsection{Results in the massless case} \label{sec:result_massless}

In the massless case we encounter a huge simplification, so there is a good reason to consider it separately. Firstly, amplitudes with fermions of the same chirality vanish:
\begin{equation}
    A_{n, +-}^\mu = 0 = A_{n, -+}^\mu.
\end{equation}
The nonzero amplitudes are given as:
\begin{equation}
    A_{n, ++}^\mu = (v+a) R_{n, ++}^\mu \quad \text{and} \quad A_{n, --}^\mu = (v-a) L_{n, --}^\mu.
\end{equation}
Below we give formulas only for $R_{n, +-}$. Other amplitudes may be reconstructed from the following symmetry:
\begin{equation}
    L_{n, --}^\mu = \overline{R}_{n, ++}^\mu,
\end{equation}
which is the consequence of the symmetries \eqref{symmetry3} and \eqref{symmetry4}

We obtained the following results for $R_{n, +-}^\mu$:
\begin{equation}
\begin{split}
    R_{1, ++}^{\dot A A} =& \asmel{p_3}{\gamma^{\dot A A} \frac{\widehat{v}_1}{v_1^2} \widehat{P}_1 \frac{\widehat{v}_2}{v_2^2} \widehat{P}_2}{-p_4} =- \frac{2S}{v_1^2 v_2^2 \sqrt{p_3^+ p_4^+}} p_4^{\dot 0 0} v_2^{\dot 0 1} p_3^{\dot A 0} v_1^{\dot 1 A}, \\
    R_{2, ++}^{\dot A A} =& \asmel{p_3}{\widehat{P}_1 \frac{\widehat{v}_3}{v_3^2} \gamma^{\dot A A} \frac{\widehat{v}_2}{v_2^2} \widehat{P}_2}{-p_4} =- \frac{2S}{v_3^2 v_2^2 \sqrt{p_3^+ p_4^+}} p_4^{\dot 0 0} p_3^{\dot 1 0} v_3^{\dot A 1} v_2^{\dot 0 A}, \\
    R_{3, ++}^{\dot A A} =& \asmel{p_3}{\widehat{P}_1 \frac{\widehat{v}_3}{v_3^2} \widehat{P}_2 \frac{\widehat{v}_4}{v_4^2} \gamma^{\dot A A}}{-p_4} =- \frac{2S}{v_3^2 v_4^2 \sqrt{p_3^+ p_4^+}} p_3^{\dot 1 0} v_3^{\dot 0 1} v_4^{\dot A 0} p_4^{\dot 0 A}, \\
    R_{4, ++}^{\dot A A} =& \asmel{p_3}{\gamma^{\dot A A} \frac{\widehat{v}_1}{v_1^2} \widehat{P}_2 \frac{\widehat{v}_5}{v_5^2} \widehat{P}_1}{-p_4} =- \frac{2S}{v_1^2 v_5^2 \sqrt{p_3^+ p_4^+}} p_4^{\dot 0 1} v_5^{\dot 1 0} p_3^{\dot A 0} v_1^{\dot 0 A}, \\
    R_{5, ++}^{\dot A A} =& \asmel{p_3}{\widehat{P}_2 \frac{\widehat{v}_6}{v_6^2} \gamma^{\dot A A} \frac{\widehat{v}_5}{v_5^2} \widehat{P}_1}{-p_4} =- \frac{2S}{v_6^2 v_5^2 \sqrt{p_3^+ p_4^+}} p_3^{\dot 0 0} p_4^{\dot 0 1} v_6^{\dot A 0} v_5^{\dot 1 A}, \\
    R_{6, ++}^{\dot A A} =& \asmel{p_3}{\widehat{P}_2 \frac{\widehat{v}_6}{v_6^2} \widehat{P}_1 \frac{\widehat{v}_4}{v_4^2} \gamma^{\dot A A}}{-p_4} = -\frac{2S}{v_6^2 v_4^2 \sqrt{p_3^+ p_4^+}} p_3^{\dot 0 0} v_6^{\dot 1 0} v_4^{\dot A 1} p_4^{\dot 0 A}, \\
    R_{7, ++}^{\dot A A} =& \frac{1}{(k_1 + k_2)^2} \asmel{p_3}{\gamma^{\dot A A} \frac{\widehat{v}_1}{v_1^2} \widehat{V}_{\mathrm{eff}}}{-p_4} = -\frac{2}{v_1^2 (k_1 + k_2)^2 \sqrt{p_3^+ p_4^+}} p_3^{\dot A 0} p_4^{\dot 0 B} \left( V_{\mathrm{eff}} \right)_{B \dot B} v_1^{\dot B A}, \\
    R_{8, ++}^{\dot A A} =& \frac{1}{(k_1 + k_2)^2} \asmel{p_3}{\widehat{V}_{\mathrm{eff}} \frac{\widehat{v}_4}{v_4^2} \gamma^{\dot A A}}{-p_4} = -\frac{2}{v_4^2 (k_1 + k_2)^2 \sqrt{p_3^+ p_4^+}} v_4^{\dot A B} \left( V_{\mathrm{eff}} \right)_{B \dot B} p_3^{\dot B 0} p_4^{\dot 0 A}.
\end{split}
\end{equation}

These results take the simplest forms in terms of the spinor notation, but as an example we give two formulas written in Lorentz vector notation:

\begin{equation}
    R_{1, ++}^\mu = -\frac{2S}{v_1^2 v_2^2} \sqrt{\frac{p_4^+}{p_3^+}} v_2^{\overline{\perp}} \begin{pmatrix}
        p_3^+ v_1^\perp & p_3^\perp v_1^- & p_3^\perp v_1^\perp & p_3^+ v_1^-
    \end{pmatrix}^\mu
\end{equation}
\begin{equation}
    R_{7, ++}^\mu = -\frac{4}{v_1^2 (k_1 + k_2)^2 \sqrt{p_3^+ p_4^+}} M_7^{\mu \nu} \left( V_{\mathrm{eff}} \right)_\nu, \\
\end{equation}
where $M_7$ is the following matrix:
\begin{equation}
    M_7^{\mu \nu} = \begin{pmatrix}
        p_3^+ p_4^+ v_1^+ & p_3^+ p_4^{\overline{\perp}} v_1^\perp & p_3^+ p_4^+ v_1^\perp & p_3^+ p_4^{\overline{\perp}} v_1^+ \\
        p_3^\perp p_4^+ v_1^{\overline{\perp}} & p_3^\perp p_4^{\overline{\perp}} v_1^- & p_3^\perp p_4^+ v_1^- & p_3^\perp p_4^{\overline{\perp}} v_1^{\overline{\perp}} \\
        p_3^\perp p_4^+ v_1^+ & p_3^\perp p_4^{\overline{\perp}} v_1^\perp & p_3^\perp p_4^+ v_1^\perp & p_3^\perp p_4^{\overline{\perp}} v_1^+ \\
        p_3^+ p_4^+ v_1^{\overline{\perp}} & p_3^+ p_4^{\overline{\perp}} v_1^- & p_3^+ p_4^+ v_1^- & p_3^+ p_4^{\overline{\perp}} v_1^{\overline{\perp}} \\
    \end{pmatrix}^{\mu \nu}.
\end{equation}
We remark that the simple structure of amplitudes written in terms of spinors is reflected in the fact that $M_7$ is a matrix of rank at most $2$.

\subsection{General results} \label{sec:result_massive}

In the massive case we have $8$ expressions to compute for each diagram: left and right amplitudes with two possible values of spin for the quark and for the antiquark. In~contrast to the massless case none of these expressions vanish identically, but we can still reduce the number of independent amplitudes by a factor of two using the symmetries
\begin{equation}
R_{n, - +}^\mu = \overline{L}_{n,+-}^\mu, \quad L_{n, - +}^\mu = \overline{R}_{n,+-}^\mu, \quad R_{n, - -}^\mu =- \overline{L}_{n,++}^\mu, \quad L_{n, - -}^\mu = -\overline{R}_{n,++}^\mu,
\end{equation}
which, similarly like in massless case, are reflection of the symmetries \eqref{symmetry3} and \eqref{symmetry4}. Therefore for each diagram we write only $R_{n, ++}^\mu, L_{n, ++}^\mu, R_{n, +-}^\mu$ and $L_{n, +-}^\mu$.

The first diagram:
\begin{equation}
\begin{split}
    R_{1, ++}^{\dot A A} =& -\frac{2S}{\left( v_1^2 - m_4^2 \right) \left( v_2^2 - m_4^2 \right)} \frac{p_4^{\dot 0 0} p_3^{\dot A 0}}{\sqrt{p_3^+ p_4^+}} \left[ v_2^{\dot 0 1} v_1^{\dot 1 A} + m_4^2 o^A \right], \\
    L_{1, ++}^{\dot A A} =& -\frac{2S}{\left( v_1^2 - m_4^2 \right) \left( v_2^2 - m_4^2 \right)} \frac{m_3 m_4 p_4^{\dot 0 0}}{\sqrt{p_3^+ p_4^+}} \left[ v_1^{\dot A 1} + v_2^{\dot 0 1} \overline{o}^{\dot A} \right] \iota^A,
\end{split}
\end{equation}
\begin{equation}
\begin{split}
    R_{1, +-}^{\dot A A} =& -\frac{2S}{\left( v_1^2 - m_4^2 \right) \left( v_2^2 - m_4^2 \right)} \frac{m_4 p_4^{\dot 0 0} p_3^{\dot A 0}}{\sqrt{p_3^+ p_4^+}} \left[ v_2^{\dot 1 0} o^A - v_1^{\dot 1 A} \right], \\
    L_{1, +-}^{\dot A A} =& -\frac{2S}{\left( v_1^2 - m_4^2 \right) \left( v_2^2 - m_4^2 \right)} \frac{m_3 p_4^{\dot 0 0}}{\sqrt{p_3^+ p_4^+}} \left[ v_2^{\dot 1 0} v_1^{\dot A 1} + m_4^2 \overline{o}^{\dot A} \right] \iota^A,
\end{split}
\end{equation}

The second diagram:
\begin{equation}
\begin{split}
    R_{2, ++}^{\dot A A} =& -\frac{2S}{\left( v_3^2 - m_3^2 \right) \left( v_2^2 - m_4^2 \right)} \frac{p_4^{\dot 0 0}}{\sqrt{p_3^+ p_4^+}} \left[ p_3^{\dot 1 0} v_3^{\dot A 1} + m_3^2 \overline{o}^{\dot A} \right] v_2^{\dot 0 A}, \\
    L_{2, ++}^{\dot A A} =& -\frac{2S}{\left( v_3^2 - m_3^2 \right) \left( v_2^2 - m_4^2 \right)} \frac{m_3 m_4 p_4^{\dot 0 0}}{\sqrt{p_3^+ p_4^+}} \left[ v_3^{\dot A 1} + v_2^{\dot 0 1} \overline{o}^{\dot A} \right] \overline{\iota}^{\dot A},
\end{split}
\end{equation}
\begin{equation}
\begin{split}
    R_{2, +-}^{\dot A A} =& \frac{2S}{\left( v_3^2 - m_3^2 \right) \left( v_2^2 - m_4^2 \right)} \frac{m_4 p_4^{\dot 0 0}}{\sqrt{p_3^+ p_4^+}} \left[ p_3^{\dot 1 0} v_3^{\dot A 1} + m_3^2 \overline{o}^{\dot A} \right] \iota^A, \\
    L_{2, +-}^{\dot A A} =& -\frac{2S}{\left( v_3^2 - m_3^2 \right) \left( v_2^2 - m_4^2 \right)} \frac{m_3 p_4^{\dot 0 0}}{\sqrt{p_3^+ p_4^+}} \left[ v_3^{\dot 1 A} - p_3^{\dot 1 0} o^A \right] v_2^{\dot A 0},
\end{split}
\end{equation}

The third diagram:
\begin{equation}
\begin{split}
    R_{3, ++}^{\dot A A} =& -\frac{2S}{\left( v_3^2 - m_3^2 \right) \left( v_4^2 - m_3^2 \right)} \frac{p_4^{\dot 0 A}}{\sqrt{p_3^+ p_4^+}} \left[ p_3^{\dot 1 0} v_3^{\dot 0 1} v_4^{\dot A 0} + m_3^2 \left( (p_3 - v_3)^{\dot 1 0} \overline{\iota}^{\dot A} + v_4^{\dot A 0} \right) \right] v_2^{\dot 0 A}, \\
    L_{3, ++}^{\dot A A} =& \frac{2S}{\left( v_3^2 - m_3^2 \right) \left( v_4^2 - m_3^2 \right)} \frac{m_3 m_4}{\sqrt{p_3^+ p_4^+}} \left[ \left( m_3^2 + p_3^{\dot 1 0} v_3^{\dot 0 1} \right) \iota^A - (p_3 - v_3)^{\dot 1 0} v_4^{\dot 0 A} \right] \overline{\iota}^{\dot A},
\end{split}
\end{equation}
\begin{equation}
\begin{split}
    R_{3, +-}^{\dot A A} =& -\frac{2S}{\left( v_3^2 - m_3^2 \right) \left( v_4^2 - m_3^2 \right)} \frac{m_4}{\sqrt{p_3^+ p_4^+}} \left[ \left( m_3^2 + p_3^{\dot 1 0} v_3^{\dot 0 1} \right) v_4^{\dot A 1} + m_3^2 (p_3 - v_3)^{\dot 1 0} \overline{\iota}^{\dot A} \right] \iota^A, \\
    L_{3, +-}^{\dot A A} =& -\frac{2S}{\left( v_3^2 - m_3^2 \right) \left( v_4^2 - m_3^2 \right)} \frac{m_3 p_4^{\dot 0 0}}{\sqrt{p_3^+ p_4^+}} \left[ \left( m_3^2 + p_3^{\dot 1 0} v_3^{\dot 0 1} \right) \iota^A - (p_3 - v_3)^{\dot 1 0} v_4^{\dot 0 A} \right],
\end{split}
\end{equation}

The fourth diagram:
\begin{equation}
\begin{split}
    R_{4, ++}^{\dot A A} =& \frac{2S}{\left( v_1^2 - m_4^2 \right) \left( v_5^2 - m_4^2 \right)} \frac{p_3^{\dot A 0}}{\sqrt{p_3^+ p_4^+}} \left[ \left( m_4^2 - p_4^{\dot 0 1} v_5^{\dot 1 0} \right) v_1^{\dot A 0} - m_4^2 (p_4 + v_5)^{\dot 0 1} \iota^A \right], \\
    L_{4, ++}^{\dot A A} =& \frac{2S}{\left( v_1^2 - m_4^2 \right) \left( v_5^2 - m_4^2 \right)} \frac{m_3 m_4}{\sqrt{p_3^+ p_4^+}} \left[ \left( m_4^2 - p_4^{\dot 0 1} v_5^{\dot 1 0} \right) \overline{\iota}^{\dot A} + (p_4 + v_5)^{\dot 0 1} v_1^{\dot A 0} \right] \iota^A,
\end{split}
\end{equation}
\begin{equation}
\begin{split}
    R_{4, +-}^{\dot A A} =& -\frac{2S}{\left( v_1^2 - m_4^2 \right) \left( v_5^2 - m_4^2 \right)} \frac{m_4 p_3^{\dot A 0}}{\sqrt{p_3^+ p_4^+}} \left[ \left( m_4^2 - p_4^{\dot 1 0} v_5^{\dot 0 1} \right) \iota^A + m_3^2 (p_4 + v_5)^{\dot 1 0} v_1^{\dot 0 A} \right], \\
    L_{4, +-}^{\dot A A} =& \frac{2S}{\left( v_1^2 - m_4^2 \right) \left( v_5^2 - m_4^2 \right)} \frac{m_3}{\sqrt{p_3^+ p_4^+}} \left[ \left( m_4^2 - p_4^{\dot 1 0} v_5^{\dot 0 1} \right) v_1^{\dot A 0} - (p_4 + v_5)^{\dot 1 0} \overline{\iota}^{\dot A} \right] \iota^A,
\end{split}
\end{equation}

The fifth diagram:
\begin{equation}
\begin{split}
    R_{5, ++}^{\dot A A} =& \frac{2S}{\left( v_6^2 - m_3^2 \right) \left( v_5^2 - m_4^2 \right)} \frac{p_3^{\dot 0 0}}{\sqrt{p_3^+ p_4^+}} \left[ m_4^2 o^A - p_4^{\dot 0 1} v_5^{\dot 1 A} \right] v_6^{\dot A 0}, \\
    L_{5, ++}^{\dot A A} =& \frac{2S}{\left( v_6^2 - m_3^2 \right) \left( v_5^2 - m_4^2 \right)} \frac{m_3 m_4 p_3^{\dot 0 0}}{\sqrt{p_3^+ p_4^+}} \left[ v_5^{\dot A 1} + p_4^{\dot 0 1} \overline{o}^{\dot A} \right] \iota^A,
\end{split}
\end{equation}
\begin{equation}
\begin{split}
    R_{5, +-}^{\dot A A} =& -\frac{2S}{\left( v_6^2 - m_3^2 \right) \left( v_5^2 - m_4^2 \right)} \frac{m_4 p_3^{\dot 0 0}}{\sqrt{p_3^+ p_4^+}} \left[ v_5^{\dot 1 A} + p_4^{\dot 1 0} o^A \right] v_6^{\dot A 0}, \\
    L_{5, +-}^{\dot A A} =& -\frac{2S}{\left( v_6^2 - m_3^2 \right) \left( v_5^2 - m_4^2 \right)} \frac{m_3 p_3^{\dot 0 0}}{\sqrt{p_3^+ p_4^+}} \left[ m_4^2 \overline{o}^{\dot A} - p_4^{\dot 1 0} v_5^{\dot A 1} \right] \iota^A,
\end{split}
\end{equation}

The sixth diagram:
\begin{equation}
\begin{split}
    R_{6, ++}^{\dot A A} =& -\frac{2S}{\left( v_6^2 - m_3^2 \right) \left( v_4^2 - m_3^2 \right)} \frac{p_3^{\dot 0 0} p_4^{\dot 0 A}}{\sqrt{p_3^+ p_4^+}} \left[ v_6^{\dot 1 0} v_4^{\dot A 1} + m_3^2 \overline{o}^{\dot A} \right], \\
    L_{6, ++}^{\dot A A} =& \frac{2S}{\left( v_6^2 - m_3^2 \right) \left( v_4^2 - m_3^2 \right)} \frac{m_3 m_4 p_3^{\dot 0 0}}{\sqrt{p_3^+ p_4^+}} \left[ v_4^{\dot 1 A} - v_6^{\dot 1 0} o^A \right] \overline{\iota}^{\dot A},
\end{split}
\end{equation}
\begin{equation}
\begin{split}
    R_{6, +-}^{\dot A A} =& -\frac{2S}{\left( v_6^2 - m_3^2 \right) \left( v_4^2 - m_3^2 \right)} \frac{m_4 p_3^{\dot 0 0}}{\sqrt{p_3^+ p_4^+}} \left[ v_4^{\dot A 1} v_6^{\dot 1 0} + m_3^2 \overline{o}^{\dot A} \right] \iota^A, \\
    L_{6, +-}^{\dot A A} =& \frac{2S}{\left( v_6^2 - m_3^2 \right) \left( v_4^2 - m_3^2 \right)} \frac{m_3 p_3^{\dot 0 0} p_4^{\dot A 0}}{\sqrt{p_3^+ p_4^+}} \left[ v_6^{\dot 1 0} o^A - v_4^{\dot 1 A} \right],
\end{split}
\end{equation}

The seventh diagram
\begin{equation}
\begin{split}
R_{7, ++}^{\dot A A} =& \frac{2}{\left(v_1^2 - m_4^2\right) (k_1 + k_2)^2} \frac{p_3^{\dot A 0}}{\sqrt{p_3^+ p_4^+}} \left[ -p_4^{\dot{0} B} \left( V_{\mathrm{eff}} \right)_{B \dot B} v_1^{\dot B A} + m_4^2 \left( V_{\mathrm{eff}} \right)^{\dot 0 A} \right], \\
L_{7, ++}^{\dot A A} =& -\frac{2}{\left(v_1^2 - m_4^2\right) (k_1 + k_2)^2} \frac{m_3 m_4 \varepsilon_{BC}}{\sqrt{p_3^+ p_4^+}} \left[ v_1^{\dot A B} \left( V_{\mathrm{eff}} \right)^{\dot 0 C} - \left( V_{\mathrm{eff}} \right)^{\dot A B} p_4^{\dot 0 C} \right] \iota^A,
\end{split}
\end{equation}
\begin{equation}
\begin{split}
R_{7, +-}^{\dot A A} =& \frac{2}{\left(v_1^2 - m_4^2\right) (k_1 + k_2)^2} \frac{m_4 p_3^{\dot A 0} \varepsilon_{\dot B \dot C}}{\sqrt{p_3^+ p_4^+}} \left[ v_1^{\dot B A} \left( V_{\mathrm{eff}} \right)^{\dot C 0} - \left( V_{\mathrm{eff}} \right)^{\dot B A} p_4^{\dot C 0} \right], \\
L_{7, +-}^{\dot A A} =& \frac{2}{\left(v_1^2 - m_4^2\right) (k_1 + k_2)^2} \frac{m_3}{\sqrt{p_3^+ p_4^+}} \left[ -v_1^{\dot A B} \left( V_{\mathrm{eff}} \right)_{B \dot B} p_4^{\dot B 0} + m_4^2 \left( V_{\mathrm{eff}} \right)^{\dot A 0} \right] \iota^A,
\end{split}
\end{equation}

The eighth diagram:
\begin{equation}
\begin{split}
R_{8, ++}^{\dot A A} =& -\frac{2}{\left(v_4^2 - m_3^2\right) (k_1 + k_2)^2} \frac{p_4^{\dot 0 A}}{\sqrt{p_3^+ p_4^+}} \left[ v_4^{\dot A B} \left( V_{\mathrm{eff}} \right)_{B \dot B} p_3^{\dot B 0} + m_3^2 \left( V_{\mathrm{eff}} \right)^{\dot A 0} \right], \\
L_{8, ++}^{\dot A A} =& -\frac{2}{\left(v_4^2 - m_3^2 \right) (k_1 + k_2)^2} \frac{m_3 m_4 \varepsilon_{\dot B \dot C}}{\sqrt{p_3^+ p_4^+}} \left[ v_4^{\dot B A} \left( V_{\mathrm{eff}} \right)^{\dot C 0} + \left( V_{\mathrm{eff}} \right)^{\dot B A} p_3^{\dot C 0} \right] \overline{\iota}^{\dot A},
\end{split}
\end{equation}
\begin{equation}
\begin{split}
R_{8, +-}^{\dot A A} =& -\frac{2}{\left(v_4^2 - m_3^2\right) (k_1 + k_2)^2} \frac{m_4}{\sqrt{p_3^+ p_4^+}} \left[ v_4^{\dot A B} \left( V_{\mathrm{eff}} \right)_{B \dot B} p_3^{\dot B 0} + m_3^2 \left( V_{\mathrm{eff}} \right)^{\dot A 0} \right] \iota^A, \\
L_{8, +-}^{\dot A A} =& \frac{2}{\left(v_4^2 - m_3^2\right) (k_1 + k_2)^2} \frac{m_3 p_4^{\dot A 0} \varepsilon_{\dot B \dot C}}{\sqrt{p_3^+ p_4^+}} \left[ v_4^{\dot B A} \left( V_{\mathrm{eff}} \right)^{\dot C 0} + \left( V_{\mathrm{eff}} \right)^{\dot B A} p_3^{\dot C 0} \right].
\end{split}
\end{equation}


\subsection{Calculations} \label{sec:calc}

In order to illustrate our calculation methods, we will now compute the amplitude $R_{1, ++}^\mu$ for both the massless and massive case.

In order to do that, we can use the fact that $P_1 = \sqrt{S} l_+$, $P_2 = \sqrt{S} l_-$, the eigenequations \eqref{eq:spinor_chirality} and the completeness relation \eqref{completeness relation for basis}:
\begin{equation}
\begin{split}
    \frac{v_1^2 v_2^2}{2S} R_{1, ++}^\mu =& \frac{1}{2} \asmel{p_3}{P_+ \gamma^\mu \widehat{v}_1 \widehat{l}_+ \widehat{v}_2 \widehat{l}_-}{-p_4} = \\
    =& \frac{1}{2} \asmel{p_3}{\gamma^\mu \widehat{v}_1 \left( \ket{\uparrow}\sbra{\uparrow} + \sket{\uparrow}\bra{\uparrow} \right) \widehat{v}_2 \left( \ket{\downarrow}\sbra{\downarrow} + \sket{\downarrow}\bra{\downarrow} \right)}{-p_4} = \\
    =& \frac{1}{2} \mel{p_3}{\gamma^\mu \widehat{v}_1}{\uparrow} \samel{\uparrow}{\widehat{v}_2}{\downarrow} \sbraket{\downarrow}{-p_4}.
\end{split}
\label{label1234}
\end{equation}
With the help of \eqref{completeness relation for basis} and \eqref{bispinors in basis} we evaluate:
\begin{equation}
    \samel{\uparrow}{\widehat{v}_2}{\downarrow} = -v_2^{\overline{\perp}}, \qquad  \sbraket{\downarrow}{-p_4}=-\sqrt{p_4^+}.
\end{equation}
Next we will insert the identity between $\gamma^\mu$ and $\widehat{v}_1$ using the completeness relation \eqref{Completeness relation} to obtain
\begin{equation}
\begin{split}
    \eqref{label1234} =& \frac{v_2^{\overline{\perp}} \sqrt{p_4^+}}{2} \left( \asmel{p_3}{\gamma^\mu}{\uparrow} \samel{\downarrow}{\widehat{v}_1}{\uparrow} - \asmel{p_3}{\gamma^\mu}{\downarrow} \samel{\uparrow}{\widehat{v}_1}{\uparrow} \right).
\end{split}
\label{label567}
\end{equation}
Then again from \eqref{completeness relation for basis} we get $\samel{\downarrow}{\widehat{v}_1}{\uparrow} = -v_1^{\perp}$, $\samel{\uparrow}{\widehat{v}_1}{\uparrow} = v_1^-$, and also we decompose $\bra{p_3}$ with \eqref{bispinors in basis}. These procedures leads to
\begin{equation}
\begin{split}
    \eqref{label567} =& -\frac{v_2^{\overline{\perp}}}{2} \sqrt{\frac{p_4^+}{p_3^+}} \Big[ v_1^\perp \left( p_3^+ \asmel{\uparrow}{\gamma^\mu}{\uparrow} + p_3^\perp \asmel{\downarrow}{\gamma^\mu}{\uparrow} \right) + \\
    &\hspace{2cm} + v_1^- \left( p_3^+ \asmel{\uparrow}{\gamma^\mu}{\downarrow} + p_3^\perp \asmel{\downarrow}{\gamma^\mu}{\downarrow} \right) \Big] = \\
    =& -v_2^{\overline{\perp}} \sqrt{\frac{p_4^+}{p_3^+}} \myvec{v_1^\perp p_3^+ & v_1^- p_3^\perp & v_1^\perp p_3^\perp & v_1^- p_3^+ }^\mu.
\end{split}
\end{equation}

We can also calculate the amplitude using the abstract index notation in spinor form
\begin{equation}
\begin{split}
    \frac{v_1^2 v_2^2 \sqrt{p_3^+ p_4^+}}{2S} R_{1, ++}^{\dot{A}A} =& \frac{\sqrt{p_3^+ p_4^+}}{2 S} \asmel{p_3}{P_+ \gamma^{\dot{A}A} \widehat{v}_1 \widehat{P}_1 \widehat{v}_2 \widehat{P}_2}{-p_4} = \\
    =& \begin{bmatrix} 0 & (p_3)_{1 \dot{B}} \end{bmatrix} \begin{bmatrix} 0 & 0 \\ \varepsilon^{\dot{A} \dot{B}} \varepsilon^{AB} & 0 \end{bmatrix} \begin{bmatrix} 0 & (v_1)_{B \dot{C}} \\ v_1^{\dot{B} C} & 0 \end{bmatrix} \begin{bmatrix} 0 & o_C \overline{o}_{\dot{C}} \\ o^C \overline{o}^{\dot{C}} & 0 \end{bmatrix} \times \\
    &\hspace{2.2cm} \times \begin{bmatrix} 0 & (v_2)_{C \dot{D}} \\ v_2^{\dot{C} D} & 0 \end{bmatrix} \begin{bmatrix} 0 & \iota_D \overline{\iota}_{\dot{D}} \\ \iota^D \overline{\iota}^{\dot{D}} & 0 \end{bmatrix} \begin{bmatrix} -(p_4)_{D \dot{1}} \\ 0 \end{bmatrix} = \\
    =& -(p_3)_{1\dot{B}} \varepsilon^{\dot{A} \dot{B}} \varepsilon^{AB} (v_1)_{B \dot{C}} \overline{o}^{\dot{C}} o^C (v_2)_{C \dot{D}} \iota^D \overline{\iota}^{\dot{D}} (p_4)_{D \dot{1}} = \\
    =& -\tensor{(p_3)}{_1^{\dot{A}}} \tensor{(v_1)}{^A_{\dot{0}}} (v_2)_{0 \dot{1}} (p_4)_{1 \dot{1}} = -p_4^{\dot{0} 0} v_2^{\dot{0} 1} p_3^{\dot{A} 0} v_1^{\dot{1} A}.
\end{split}
\end{equation}
For the massive case it is a little bit more complicated. If we want to get the result in bracket notation in vector form, we should multiply all the terms and then use the completeness relation in each term.

In the spinor abstract index notation, the complication of the massive cases comes from the fact that now not all the matrices are antidiagonal and the bispinors contain more terms. Let us again consider the first amplitude as an example:
\begin{equation}
\begin{split}
    \frac{\sqrt{p_3^+ p_4^+}}{2 S} &\asmel{p_3}{P_+ \gamma^{\dot{A}A} (\widehat{v}_1 + m_4) \widehat{P}_1 (\widehat{v}_2 + m_4) \widehat{P}_2}{-p_4} = \\
    =& \begin{bmatrix} m_3 \iota^B & (p_3)_{1 \dot{B}} \end{bmatrix} \begin{bmatrix} 0 & 0 \\ \varepsilon^{\dot{A} \dot{B}} \varepsilon^{AB} & 0 \end{bmatrix} \begin{bmatrix} m_4 \tensor{\delta}{_B^C} & (v_1)_{B \dot{C}} \\ v_1^{\dot{B} C} & m_4 \tensor{\delta}{^{\dot B}_{\dot C}} \end{bmatrix} \begin{bmatrix} 0 & o_C \overline{o}_{\dot{C}} \\ o^C \overline{o}^{\dot{C}} & 0 \end{bmatrix} \times \\
    &\hspace{2.8cm} \times \begin{bmatrix} m_4 \tensor{\delta}{_C^D} & (v_2)_{C \dot{D}} \\ v_2^{\dot{C} D} & m_4 \tensor{\delta}{^{\dot C}_{\dot D}} \end{bmatrix} \begin{bmatrix} 0 & \iota_D \overline{\iota}_{\dot{D}} \\ \iota^D \overline{\iota}^{\dot{D}} & 0 \end{bmatrix} \begin{bmatrix} -(p_4)_{D \dot{1}} \\ m_4 \overline{\iota}^{\dot{D}} \end{bmatrix} = \\
    =& -p_4^{\dot{0}0} p_3^{\dot A 0} \left[ v_2^{\dot 0 1} v_1^{\dot 1 A} + m_4^2 o^A \right].
\end{split}
\end{equation}
To obtain the final result one has to perform the matrix multiplication and then contract and raise some spinor indices.

\subsection{Comparison with the trace method}

For the purpose of this subsection only, we denote the spinorial matrix appearing in the amplitude $A_n^\mu$ by $\mathfrak A_n^\mu$:
\begin{equation}
    A_n^\mu = \overline u_{\sigma_3}(p_3) \mathfrak A_n^\mu v_{\sigma_4}(p_4). 
\end{equation}
In an analogous way we define the matrices $\mathfrak A_S^\mu, \mathfrak A_A^\mu$. 

The amplitudes squared \eqref{Amplitude squared with spins} summed also over spins of $q \overline q $ can be computed as traces. For the symmetric part 
\begin{equation}
\begin{split}
    \sum_{\sigma_3, \sigma_4} \mathcal{M}_S^{\mu \nu} = \frac{g^4 \left( N^2-2 \right)}{2N \left( N^2 - 1 \right)} \Tr[\left( \widehat{p}_3 + m_3 \right) \mathfrak{A}_S^\mu \left( \widehat{p}_4 - m_4 \right) \mathfrak{A}_S^{\dag \nu}],
\end{split}
\label{eq:MS_trace}
\end{equation}
and for the antisymmetric
\begin{equation}
\begin{split}
    \sum_{\sigma_3, \sigma_4} \mathcal{M}_A^{\mu \nu} = \frac{g^4 N}{2\left( N^2 - 1 \right)} \Tr[\left( \widehat{p}_3 + m_3 \right) \mathfrak{A}_A^\mu \left( \widehat{p}_4 - m_4 \right) \mathfrak{A}_A^{\dag \nu}].
\end{split}
\label{eq:MA_trace}
\end{equation}
Using the trace formulas \eqref{eq:MS_trace} and \eqref{eq:MA_trace} we checked numerically the helicity structure functions of the form
\begin{equation}
    \epsilon^{(r)*}_\mu \mathcal{M}_S^{\mu \nu} \epsilon^{(r')}_\nu, \qquad \epsilon^{(r)*}_\mu \mathcal{M}_A^{\mu \nu} \epsilon^{(r')}_\nu,
\end{equation}
where $r, r' \in \{ +, -, 0 \}$ are basis polarizations of $V^*$ defined in a chosen reference frame. Up to negligible numerical errors, they agree with results obtained by directly evaluating amplitudes using our analytic formulas. The numerical checks were performed using Wolfram Mathematica \cite{Wolfram}. The used code is available upon request.

\section{Outlook} \label{sec:outlook}

We presented compact analytic forms of the $g^* g^* \to \overline{q}qV^*$ scattering amplitudes. Evaluation of these scattering amplitudes is needed as a subroutine in programs evaluating numerically Drell-Yan structure functions, and it is inefficient if performed using trace methods \cite{Motyka:2016lta}. Our result can be used to speed up numerical calculations in such studies, and it could be used in event generators \cite{vanHameren:2016kkz}. The calculation method we have used can be applied to other processes treated in the $k_T$ factorization framework at the leading order. Spinor helicity methods can also be applied in NLO calculations, e.g.\ to obtain the real radiative corrections. One loop amplitudes involving an off-shell gluon have been considered in \cite{Blanco:2022iai}, which also uses the spinor helicity formalism. 


\section*{Acknowledgements}

We would like to thank Leszek Motyka for introducing us to the topic of the Drell -- Yan process, discussions, and reading of the manuscript. This research was supported by the Polish National Science Centre (NCN) grant no. 2017/27/B/ST2/02755.

\end{document}